\documentclass[aip,pop,reprint]{revtex4-1}
\usepackage{dcolumn}
\usepackage{amsmath}
\usepackage{bm}
\usepackage{hyperref}
\usepackage{natbib}
\usepackage{amsfonts}
\usepackage{booktabs}
\usepackage{siunitx} 
\usepackage{relsize}
\usepackage{mathtools}
\usepackage[colorinlistoftodos]{todonotes}
\usepackage{physics}
\usepackage{graphicx}
\usepackage{array}
\usepackage{float}
\usepackage{svg}
\usepackage{tikz}
\usetikzlibrary{tikzmark}
\usepackage{empheq}
\usepackage{enumitem}
\makeatletter
\def\namedlabel#1#2{\begingroup
    #2%
    \def\@currentlabel{#2}%
    \phantomsection\label{#1}\endgroup
}

\makeatother
\usepackage{tcolorbox}
\usepackage{color}
\usepackage{amssymb}

\begin{document}

\title{Disorder-Induced Heating in Molecular Atmospheric Pressure Plasmas}

\author{J. LeVan} 
\author{M. D. Acciarri}
\address{Department of Nuclear Engineering and Radiological Sciences, University of Michigan, Ann Arbor, MI 48109, USA}
\author{S. D. Baalrud}
\address{Department of Nuclear Engineering and Radiological Sciences, University of Michigan, Ann Arbor, MI 48109, USA}
\email{baalrud@umich.edu}
\date{\today}

\begin{abstract}
Recent work has shown that ions are strongly coupled in atmospheric pressure plasmas when the ionization fraction is sufficiently large, leading to a temperature increase from disorder-induced heating that is not accounted for in standard modelling techniques. Here, we extend this study to molecular plasmas. 
A main finding is that the energy gained by ions in disorder-induced heating gets spread over both translational and rotational degrees of freedom on a nanosecond timescale, causing the final ion and neutral gas temperatures to be lower in the molecular case than in the atomic case. A model is developed for the equilibrium temperature that agrees well with molecular dynamics simulations. 
The model and simulations are also applied to pressures up to ten atmospheres. 
We conclude that disorder-induced heating is a significant and predictable phenomena in molecular atmospheric pressure plasmas.
\end{abstract}


\maketitle

\section{Introduction}
Cold atmospheric pressure plasmas (CAPPs) are continuously being tested and developed for a wide range of applications, including wound healing and cancer treatment \cite{10.1063/5.0008093}, water purification \cite{app11083372}, and food preservation \cite{https://doi.org/10.1111/jfpp.15070}, among others\cite{article, KUMAR2021100197, bioengineering10030280}. This comes, in large part, from their unique capacity for generating large quantities of reactive oxygen and nitrogen species (RONS). Additionally, in comparison with other plasma technologies, CAPPs offer a lower cost, complexity, and carbon footprint since they do not require the use of a vacuum chamber \cite{app11114809}. This has driven increased interest in understanding plasma dynamics, especially with concern to the transport of RONS \cite{app11114809}. A significant amount of work has already been done to characterize and model these plasmas\cite{Neyts_2014, doi:10.1080/10408398.2018.1564731, PMID:30289686, Adamovich_2022}. It is, however, also important to understand the underlying physical mechanisms that drive CAPPs to be sure they are accounted for in the models. In this paper, we examine and extend one such mechanism: disorder-induced heating (DIH), which arises from strong ion-ion coupling. 

Recent work showed that ions in atmospheric pressure plasmas are strongly coupled at ionization fractions as low as $10^{-5}$, leading to rapid ion DIH on a picosecond timescale following an instantaneous ionization\cite{Acciarri_2022}. This work modelled the effect in monatomic argon plasma and found that, at high ionization fractions ($\gtrsim 10^{-2}$), it can increase ion and neutral gas temperatures by thousands of Kelvin. In this paper, we extend the previous work to molecular plasmas and examine the effects of added rotational degrees of freedom. Since many applications of CAPPs rely on N$_2$ and O$_2$ \cite{Lotfy2019, Iuchi2018-hs, 10.1063/1.4884787}, it is important to understand how disorder-induced heating impacts molecular plasmas. We also extend the work to higher pressures (up to 10 atm) to account for the regimes seen in plasma-assisted combustion \cite{Aleksandrov11}. 

Using molecular dynamics (MD) simulations of Nitrogen gas (N$_2$), we find that DIH causes the ion translational temperature to rapidly increase over approximately one ion plasma period ($\approx$ 1 ps). This is followed by translational-rotational and ion-neutral relaxation processes that occur in roughly 1 ns. The latter process leads to fast neutral gas heating. 
A main finding is that the translational-rotational energy exchange leads to a lower total temperature resulting from DIH than in the atomic case. 
This is because the kinetic energy gained by ions from DIH gets distributed amongst more degrees of freedom in the molecular case. 
Considering N$_2$, the kinetic energy gain is eventually spread equally over 3 translational and 2 rotational degrees of freedom, for a total of 5. 
This contrasts with the 3 degrees of freedom in the atomic case. 
Furthermore, we confirm that this principle extends to molecules with 3 rotational degrees of freedom by simulating an N$_4$ molecular plasma. 

Despite the reduction in temperature due to the presence of rotational degrees of freedom, DIH can still increase the neutral gas temperature by thousands of Kelvin. We derive a model for this change in temperature using conservation of energy arguments and find that it matches well with our MD results. This model is then generalized to account for higher pressures, where DIH becomes more significant. 
Finally, we consider the common case of a gradual ionization ramp over several nanoseconds. 
As in a previous study with atomic gases, the temperature increase of ions due to DIH and the subsequent ion-neutral energy relaxation are fast compared to a typical ionization timescale in a nanosecond discharge ($\sim 10$ ns).
A consequence is that both the ion and neutral temperatures increase at the rate of ionization. 
Importantly, it is also shown that translational-rotational relaxation is fast compared to the ionization timescale - thus, in the context of fast gas heating, one must consider the effect of molecular rotation. 

Fast gas heating is a particularly important implication of DIH, with applications in flow control, combustion, and CO$_2$ disassociation \cite{Popov_2011, CHENG2022111990, Pokrovskiy_2022}. There are models for this heating based on the relaxation of excited states in atoms and molecules\cite{POPOV2022100928}. These models, however, do not consider DIH and as a consequence, may underestimate heating when the ion density is sufficiently high.  

While many atmospheric pressure plasmas do not reach high enough levels of ionization for DIH to be significant, those generated from nanosecond discharges, arc discharges, and laser ionization often do \cite{Lo_2017, Minesi_2020, vanderHorst_2012, Benilov_2008, bergeson, Bataller_2014, Bataller:19}. In these instances, we expect DIH to be an important source of fast gas heating. In laser-produced plasmas, DIH is the fastest known heating mechanism present and therefore sets the ion temperature at early times \cite{bergeson}. In nanosecond discharges, neutral gas heating predicted by DIH has shown very good agreement with time-resolved gas measurements from experimental work \cite{gradual}, giving further merit to the importance of this process.

\section{Strong Coupling and Disorder-Induced Heating}

An interaction between species $s$ and $s'$ is strongly coupled if the average potential energy is on the same order as, or greater than, the average kinetic energy. 
This can be quantified by the coupling parameter exceeding unity, $\Gamma_{ss'} \ge 1$, where 
\begin{equation}
    \Gamma_{ss'} = \frac{\phi_{ss'}(r = a_{ss'})}{k_BT} .
\end{equation}
Here, $\phi_{ss^\prime}$ is the interaction potential between $s$ and $s'$, $a_{ss'}$ is the average inter-particle spacing, and $T$ is the temperature of the two species. For particles of the same species, one can write $a_{ss} = (3/4\pi n_s)^{1/3}$, which corresponds to an average inter-particle spacing $\sim 1$ nm at atmospheric pressure.

\begin{figure*}
\centering
\includegraphics[width=\textwidth]{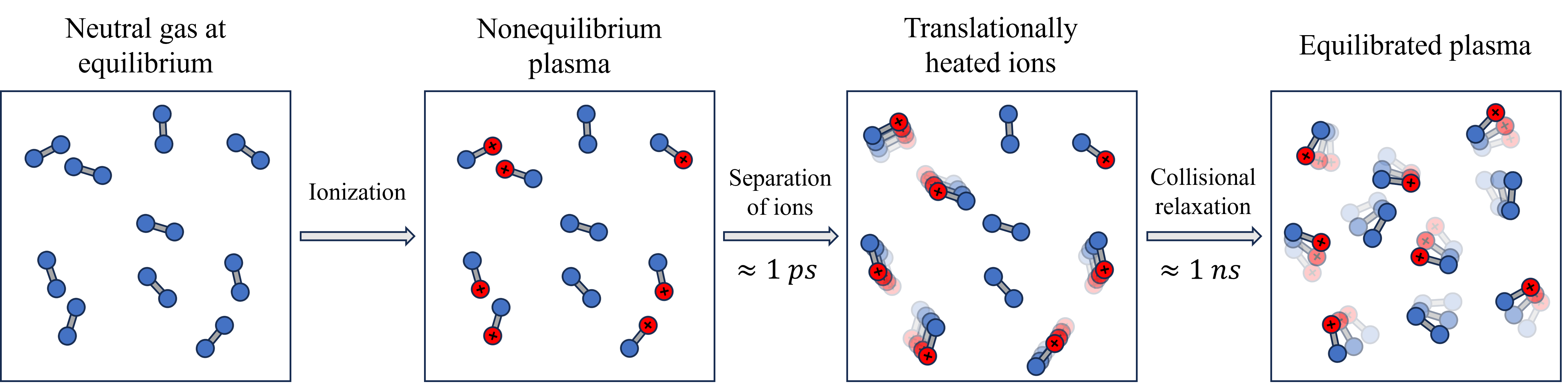}
\caption{Illustration of the disorder-induced heating process. Ionization transforms neutral gas into a nonequilibrium plasma. Coulomb repulsion causes the ions to separate on a picosecond timescale, causing significant translational heating. Collisions then distribute this energy across neutral and rotational degrees of freedom until they reach an equilibrium. The collision relaxation timescale near atmospheric pressure is often characteristic of nanoseconds.}
\label{dih}
\end{figure*}

Strongly coupled plasmas behave fundamentally differently than weakly coupled (ideal) plasmas. Binary collisions characteristic of weak coupling are replaced by many-body interactions that force the strongly coupled species into an ordered spatial configuration. Consequently, when a gas is ionized to the point of strong ion-ion coupling, the newly created ions move from a highly disordered to a more ordered state. This increased order results in a lower potential energy which, in turn, is offset by an increase in kinetic energy. This increase in kinetic energy is termed disorder-induced heating\cite{PhysRevLett.87.115003, PhysRevLett.88.065003}. The process is illustrated in figure \ref{dih}. It can be seen that the separation of ions leads to a spatial configuration with more apparent order than the initial configuration, implying a transfer of potential to kinetic energy. 

Disorder-induced heating is not a new phenomenon. It has been studied in great detail in ultracold neutral plasmas\cite{PhysRevLett.83.4776,PhysRevLett.88.065003, PhysRevE.94.021201, dihexperiment} and more recently, fusion plasmas\cite{Foster_Fetsch_Fisch_2023}. However, it has now been shown that DIH is also expected to occur in atmospheric pressure plasmas\cite{Acciarri_2022}. While this has not been confirmed experimentally, predictions from DIH have been shown to match well with experimental data \cite{Acciarri_2022, gradual}.

\section{MD Simulations}

Molecular dynamics simulations were run using the open-source software LAMMPS \cite{LAMMPS}. Nitrogen (N$_2$) molecules were modelled as rigid rotors, allowing for no vibration, and all ions were modelled as N$_2^+$, consisting of a neutral atom rigidly bonded to an ion. A bond length of 1.09 \r{A}\cite{lide2004crc}  was held constant using the ``fix rigid'' command in LAMMPS, which solves the forces and torques on each rigid body by summing the forces and torques on its constituent atoms. This approach ignores vibration, an assumption expected to be valid because the characteristic vibrational temperature of N$_2$ is $3374$ K \cite{mcquarrie1997physical} and the temperature is far below this for the majority of the simulation. Additionally, DIH and the subsequent ion-neutral energy relaxation occur at the nanosecond timescale or less, while translational-vibrational energy exchange in N$_2$ occurs on a microsecond timescale~\cite{10.1063/5.0021993}. A real ionization process will certainly create excited vibrational states, but this does not affect DIH because the ion and neutral atom temperatures are too low to couple to the vibrational modes.

Electrons were left out of the simulation because their high temperature in CAPPs make them weakly coupled. In strongly coupled plasmas, treating electrons as a non-interacting background species yields accurate ion and neutral dynamics, as commonly done in the one-component plasma (OCP) model \cite{BAUS19801}. Chemical effects were also left out as they are beyond the scope of this work. We aim not to simulate a CAPP as a whole, but instead examine a specific heating mechanism. Since DIH is caused solely by a change in the spatial orientation of ions, chemistry only affects the heating if it causes a significant change in the number density or temperature of ion species in the plasma on a relevant timescale. A model is provided (equation (\ref{eq:energy-change})) that can account for those changes. 

Interactions were modelled on a per-atom basis, with the ion-ion, ion-neutral, and neutral-neutral interaction potentials given by the Coulomb, charge-induced dipole, and Lennard-Jones potentials as follows
\begin{gather}
    \phi_{ii}(r) = \frac{q^2}{4\pi \epsilon_0}\frac{1}{r} \\
    \phi_{in}(r) = \frac{q^2}{8\pi \epsilon_0}\frac{\alpha_R a_0^3}{r^4}\left( \frac{r_\phi^8}{r^8} - 1\right) \\
    \phi_{nn}(r) = 4\epsilon\left[ \left( \frac{\sigma}{r}\right)^{12} - \left( \frac{\sigma}{r}\right)^6\right]
\end{gather}
where $\epsilon = 99.8k_B$, $\sigma = 0.3667$ nm, and $\alpha_R = 7.5$ for N$_2$ and $a_0$ is the Bohr radius. The repulsive term in the charge-induced dipole potential was added to avoid bound states, as done in previous work\cite{Acciarri_2022}. Details on this addition can be found in the Appendix of Ref.~\onlinecite{Acciarri_2022}.

Much of this paper will focus on data from the N$_2$ simulations. However, for comparison's sake, simulations of N and N$_4$ were also run. The atomic nitrogen simulations were run in the exact same fashion as the previous work\cite{Acciarri_2022}. In the N$_4$ simulations, N$_4$ was modelled using an open-chain geometry as described by Glukhovtsev and Laiter \cite{doi:10.1021/jp952026w}; a structure which, importantly, has three rotational degrees of freedom. All N$_4$ ions were modelled as N$_4^+$, with one of the two middle atoms ionized and a modified LJ coefficient $\epsilon = 29.0k_B$ was used in order to prevent significant neutral gas correlations. 

To start the simulations, 10,000 molecules of neutral gas were evolved under the influence of a thermostat until equilibrium at room temperature was reached. Then, a fraction of the molecules were instantly ionized and an NVE (microcanonical) simulation was run with timestep $\Delta t = 5\times 10^{-4}\omega_{pi}^{-1}$, a value found to be sufficiently small for conserving energy. It has been shown in previous work\cite{gradual} and is confirmed in this work (section \ref{gradual}) that the magnitude of heating from DIH is independent of ionization dynamics. 

Molecule positions and velocities were output once every plasma period and used to calculate translational and rotational temperatures. The simulation domain was a 3-dimensional box with periodic boundary conditions. 
Three simulations for each of N, N$_2$, and N$_4$ were run at every tenth ionization fraction from $0.10$ to $0.70$. Data was averaged across the three trials to minimize statistical deviation.

\section{MD Results}

\begin{figure}
     \centering
     \includegraphics[width=.5\textwidth]{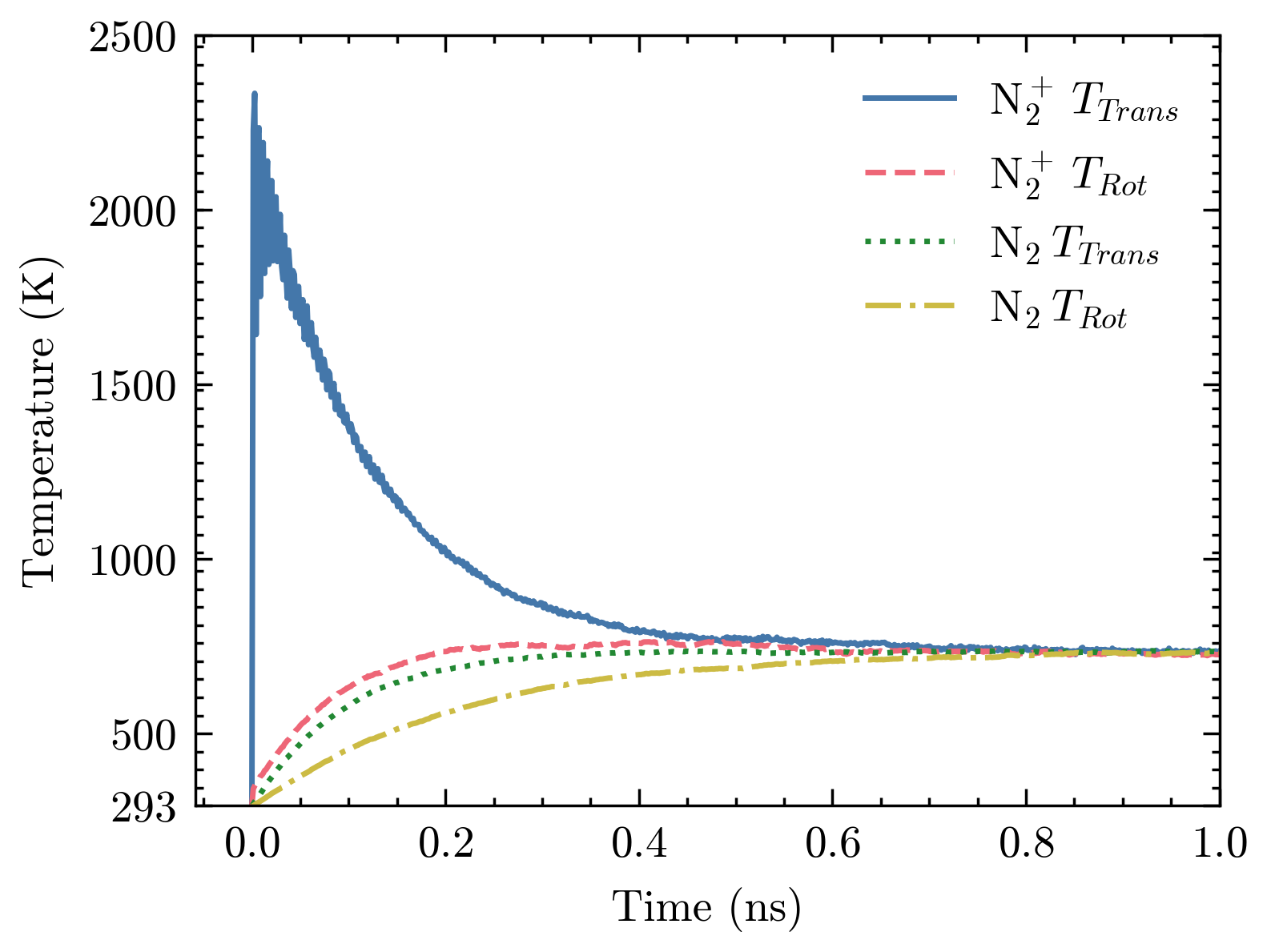}
     \caption{Simulated time-evolution of the temperature components of an N$_2$ plasma at a 0.30 ionization fraction. Lines correspond to the ion translational temperature (blue), ion rotational temperature (red), neutral molecule translational temperature (green), and neutral molecule rotational temperature (yellow).}
     \label{fig:30_ion}
 \end{figure}

The averaged time-evolution of the component temperatures from a particular set of simulations, N$_2$ at a 0.30 ionization fraction, can be seen in figure \ref{fig:30_ion}. This plot shows results of the NVE stage 
simulation and $t = 0$ refers to the time immediately following ionization. A sharp spike can be seen in the translational temperature of N$_2^+$, reaching its peak after 2.22 ps (1.52 $ \omega_{pi}^{-1}$). This is disorder-induced heating. Molecules, upon being ionized, structurally rearrange themselves until they reach their state of minimal potential energy, increasing their translational kinetic energy in the process. This heating is followed by collisional relaxation, during which the N$_2^+$ translational temperature equilibrates with the rest of the degrees of freedom in the simulation over the course of one nanosecond, ultimately reaching an equilibrated temperature of 732 K. For a clear physical picture of this process, recall figure \ref{dih}.

Simulations of each gas type, N, N$_2$, and N$_4$, followed a similar time-evolution to figure \ref{fig:30_ion}. Ion translational temperatures always peaked after 1.5 plasma periods, consistent with previous work \cite{Acciarri_2022}, and equilibration always occurred in roughly one nanosecond. However, they all produced differing equilibrium temperatures. We found the equilibrium temperature scales with the ionization fraction $x_i^{4/3}$, again consistent with the previous work\cite{Acciarri_2022}. This is caused by an increased ion density and thus an increased chance of two ions being generated in close enough proximity to gain an energy exceeding the initial thermal energy as they repel.

\begin{figure}
     \centering
     \includegraphics[width=.5\textwidth]{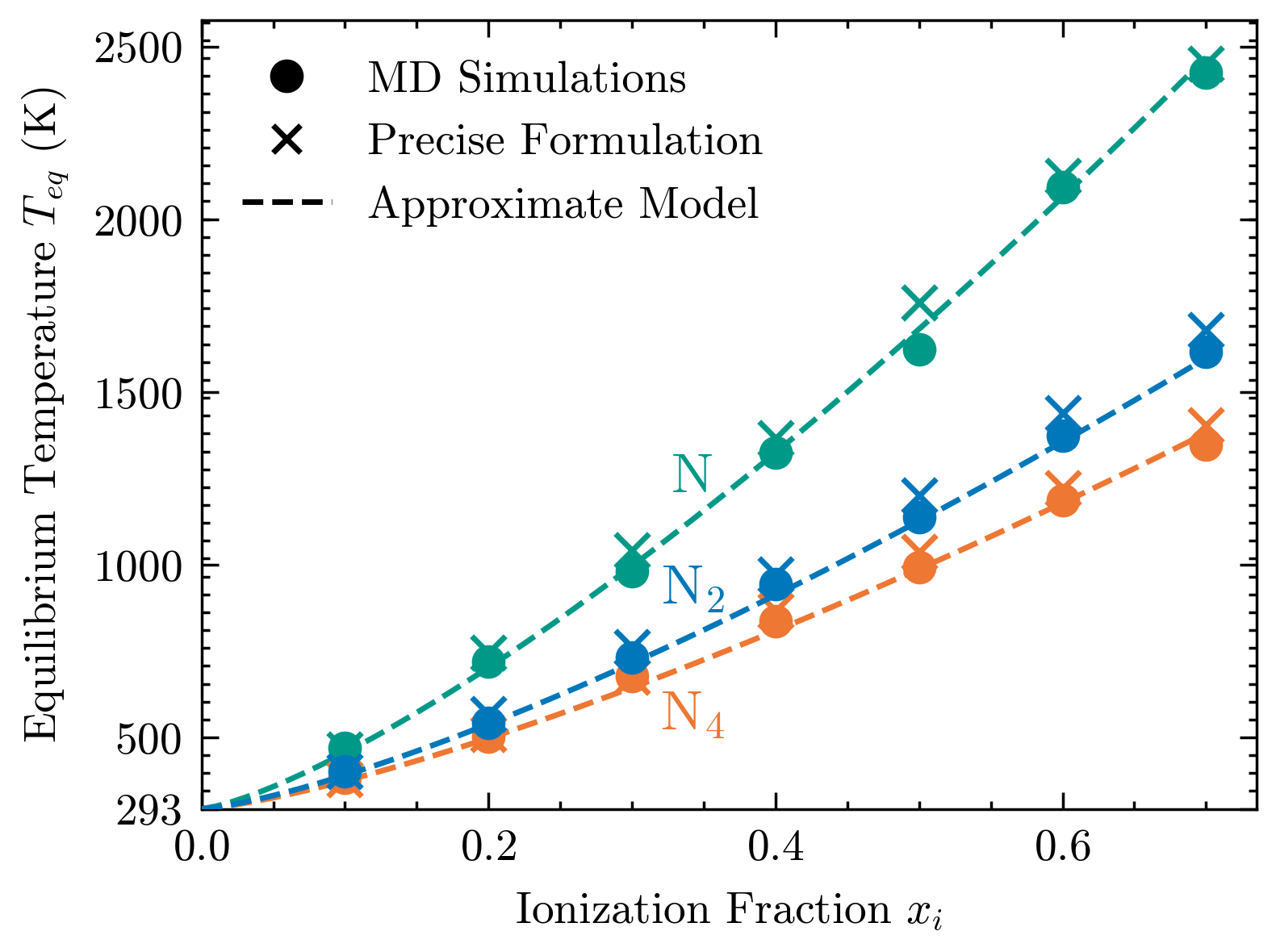}
     \caption{Simulated and predicted temperatures of N, N$_2$, and N$_4$ plasma using the precise formulation from equation~(\ref{gr-eq}) with $g_{ii}(r)$ computed from MD, and from the approximate model of equation~(\ref{eq:approximate}).}
     \label{fig:model}
 \end{figure}

\begin{figure}
     \centering
     \includegraphics[width=.5\textwidth]{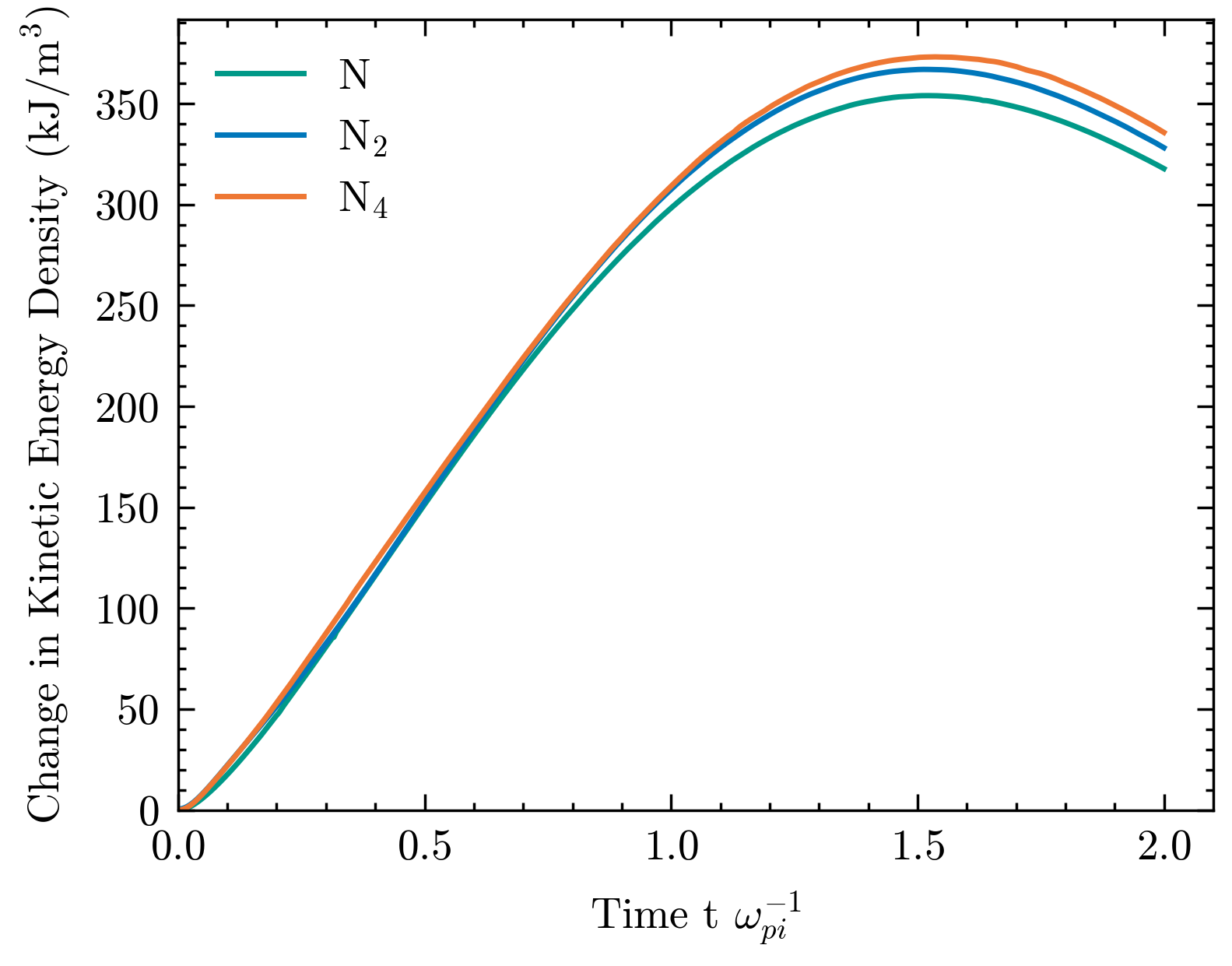}
     \caption{Change in total kinetic energy density (with respect to $t=0$) for N, N$_2$, and N$_4$. This data comes from simulations at an ionization fraction of 0.30.}
     \label{fig:ke_comp}
 \end{figure}

It is also observed that the equilibrated temperature is inversely proportional to the number of degrees of freedom present, implying $T_{eq, N} > T_{eq, N_2} > T_{eq, N_4}$, as shown in figure \ref{fig:model}. This comes as a direct consequence of the equipartition theorem. Figure \ref{fig:ke_comp} shows that, for a given ion density, N, N$_2$, and N$_4$ plasmas convert the same amount of potential energy to kinetic energy. However, in N$_4$, this energy must be spread across 6 degrees of freedom (3 translational and 3 rotational), while it is spread across just 5 in N$_2$ and 3 in N. It is clear, then, that DIH models must consider the number of active degrees of freedom.

 \section{Modelling DIH}
 \subsection{Precise Formulation}
\begin{figure}
     \centering
     \includegraphics[width=.5\textwidth]{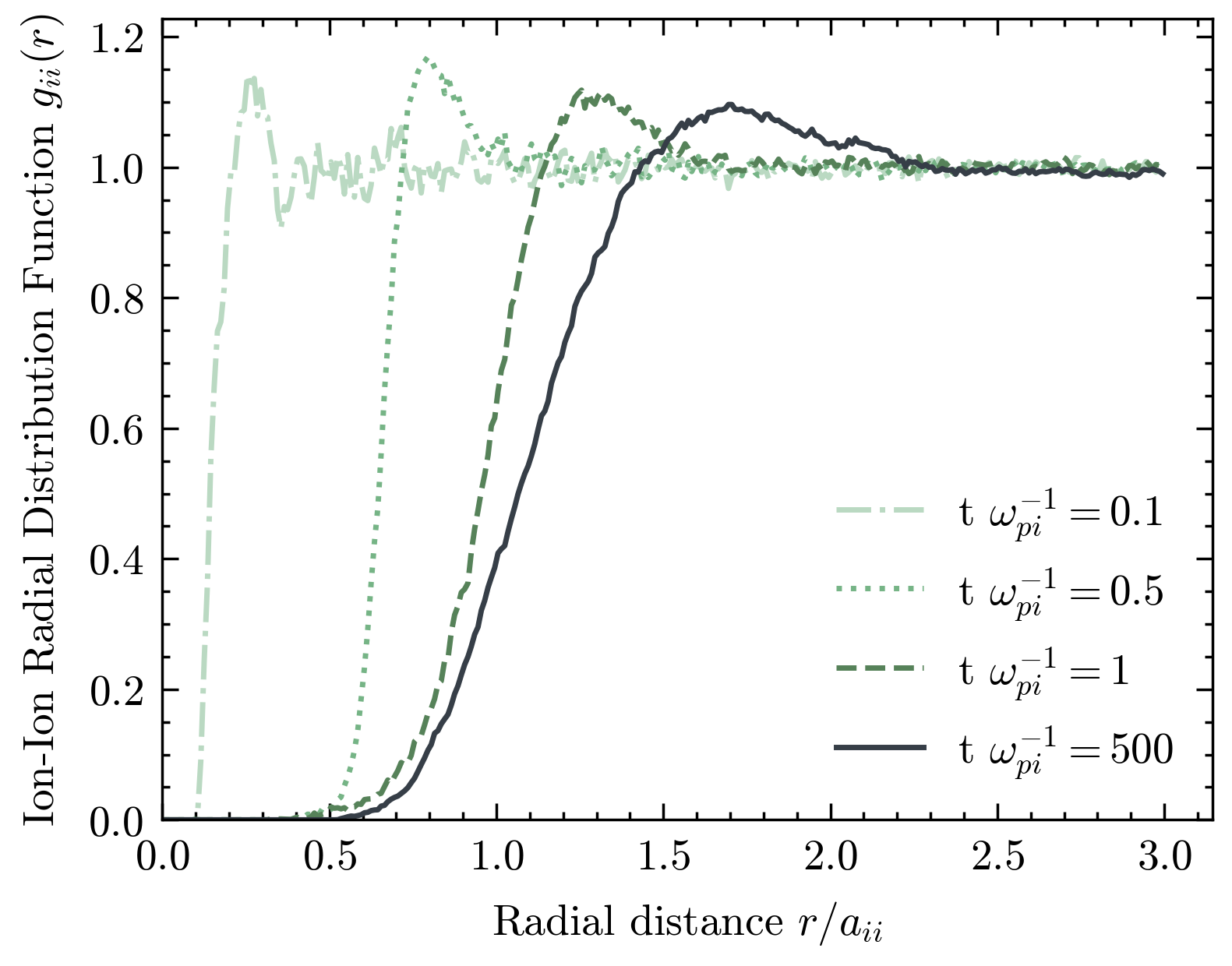}
     \caption{Snapshots of the ion-ion radial distribution function over the course of DIH from a simulation of N$_2$ at a 0.30 ionization fraction. Over time, the ion-ion radial distribution function shifts to the right as ions spread apart from each other, reducing their potential energy in the process. }
     \label{fig:rdf-over-time}
 \end{figure}

Disorder-induced heating results from a change in the spatial ordering of ions. The ion-ion radial distribution function, which outputs the ratio of the local ion density around each ion to the bulk ion density\cite{hansen2006theory}
\begin{equation}
\label{eq:gii}
    g_{ii}(r) = \frac{\langle \rho_i(r) \rangle}{\rho_i} ,
\end{equation}
 can capture this change.
Figure \ref{fig:rdf-over-time} shows an example of $g_{ii}(r)$ computed from MD simulation data using equation~(\ref{eq:gii}) at various times throughout DIH. Immediately following ionization, $g_{ii}(r) \approx 1$, indicating a disordered spatial configuration where each ion's position is effectively random. Over time, a gap forms at small radial distances, indicating the ions have spread apart from one another. This spreading apart, also depicted in figure \ref{dih}, affects the potential energy of ions.

The radial distribution function provides a convenient means to quantify the potential energy. Integrating radially and multiplying by the ion density, $\rho_i$, gives the density of ions at each distance $r + dr$ away from each reference ion. Therefore, the potential energy of ions in terms of the ion-ion distribution function is given by\cite{Gould2010-vd}
\begin{equation}
    U = \frac{N_i}{2}4 \pi \rho_i \int \phi_{ii}(r)g_{ii}(r)r^2 dr
\end{equation}
where $N_i$ is the total number of ions. 
For modelling the effect of disorder-induced heating on the equilibrium temperature, we are only concerned with how the potential energy of ions changes from immediately following ionization to equilibrium. Additionally, to avoid simulation artifacts like the number of ions, $N_i$, we are interested only in an energy density. Therefore, we convert the previous equation to a change in energy density as follows
\begin{equation}
    \Delta u_{i} = 2\pi x_i \rho_i \int \phi_{ii}(r) \left[g_{ii}(r, t_{eq}) - g_{ii}(r, t_{0})\right] r^2 dr
    \label{eq:energy-change}
\end{equation}
where $x_i$ is the ionization fraction, $\phi_{ii}(r)$ is the Coulomb potential, and $g_{ii}(r, t_{eq})$ and $g_{ii}(r, t_{0})$ are the ion-ion radial distribution functions at equilibrium and immediately following ionization respectively.

A change in the spatial orientation of ions also affects the ion-neutral interaction, which can be quantified through the ion-neutral radial distribution function ($g_{in}(r)$). However, we found 
the contribution to the potential energy change from ion-neutral interactions to be much less than from ion-ion interactions, 
and thus deemed the ion-neutral interaction to have a negligible effect on the equilibrium temperature. It is worth noting that at sufficiently high pressures, this assumption is unlikely to be valid because the ion-neutral interaction becomes strongly coupled, becoming a source of additional heating. However, at the pressures presented in this paper (up to 10 atm), we found this to be a valid approximation. 

Hence, we use equation (\ref{eq:energy-change}) to model the entirety of the change in potential energy from DIH and convert this to a temperature change using traditional thermodynamics arguments
\begin{equation}
\label{gr-eq}
    \Delta T = -\frac{2\Delta u_{i}}{f k_B},
\end{equation}
where $f$ is the number of degrees of freedom.

Figure \ref{fig:model} shows close agreement between the model and simulated temperatures. Here, the radial distribution functions used in the model were computed from the MD simulation data itself. We observe a slight discrepancy from the simulated temperatures, which can be attributed to noise in the radial distribution function. This noise can be seen in figure \ref{fig:rdf-over-time}, and comes as an artifact of the finite number of ions and finite number of timesteps over which we can compute this function. 

Equation (\ref{gr-eq}) is, by nature, a flexible formulation for the equilibrium temperature. One can consider the effects of recombination by splitting equation~(\ref{eq:energy-change}) into having an ``initial'' and ``equilibrium'' ionization fraction and ion density. Additionally, it is additive, such that if multiple species of ions exist in a plasma, one need only sum the contributions of each. To use the model, there exist several methods for estimating radial distribution functions including the Percus-Yevick\cite{STELL1963517} and hypernetted-chain\cite{10.1143/PTP.20.920} approximations. Future work will use these methods to model DIH and be implemented into a fluid model.   
This approach relies on numerically evaluating a model for the radial distribution functions. 
Next, we consider a simpler more approximate approach to estimate the potential energy change in DIH directly. 

\subsection{Approximate Model}

A simpler estimate for DIH can be constructed by analyzing the change in average ion-ion spacing. 
In the limiting case of transitioning from a completely disordered to perfectly ordered state, we expect the average ion spacing to change from approximately $a_{ii}/2$ to $a_{ii}$, leading to a change in potential energy 
\begin{equation}
\label{9}
    \Delta u_{\max} = \frac{x_ie^2}{4\pi \epsilon_0}\left(\frac{1}{a_{ii}} - \frac{1}{a_{ii}/2}\right) .
\end{equation}
We found that in the ion densities relevant to strongly coupled CAPPs, a linear correction factor of $\kappa = 2/3$ is required to fit our data, such that $\Delta u = \kappa \Delta u_{\max}$. Converting equation~(\ref{9}) to a change in temperature, we find 
\begin{equation}
    \Delta T = \frac{e^2}{2\pi \epsilon_0 k_B}\frac{ \kappa x_i}{fa_{ii}}
    \label{eq:approximate}
\end{equation}
Equation~(\ref{eq:approximate}) is a slight modification of the model used in Ref.~\onlinecite{Acciarri_2022} to account for the added degrees of freedom in molecular motion. 

A comparison of the prediction of equation~(\ref{eq:approximate}) and MD simulations is shown in figure~\ref{fig:model}. 
The close agreement with the simple model is in part due to the use of the $\kappa = 2/3$ fit coefficient, but the model captures well the scaling with ionization fraction. 
It also highlights the importance of the degrees of freedom in distributing the energy gained by DIH. The change in temperature of N ($f=3$), N$_2$ ($f=5$), and N$_4$ ($f=6$) differ only by a factor of $1/f$. While it lacks the physical precision associated with the previous model, it can still be useful as a fairly accurate means to estimate the extent to which DIH will affect a given plasma.

\section{Higher Pressures}

Simulations were also run with initial gas pressures of 2.5, 5.0, 7.5, and 10 atm at constant ionization fractions to explore the pressure dependence of DIH. This is particularly applicable to plasma-assisted combustion, where initial gas pressures can range from 0.1-10 atm\cite{Aleksandrov11} and fast gas heating is known to play a role in accelerating and intensifying the combustion process\cite{POPOV2022100928}. 

\begin{figure}
    \centering
    \includegraphics[width=0.5\textwidth]{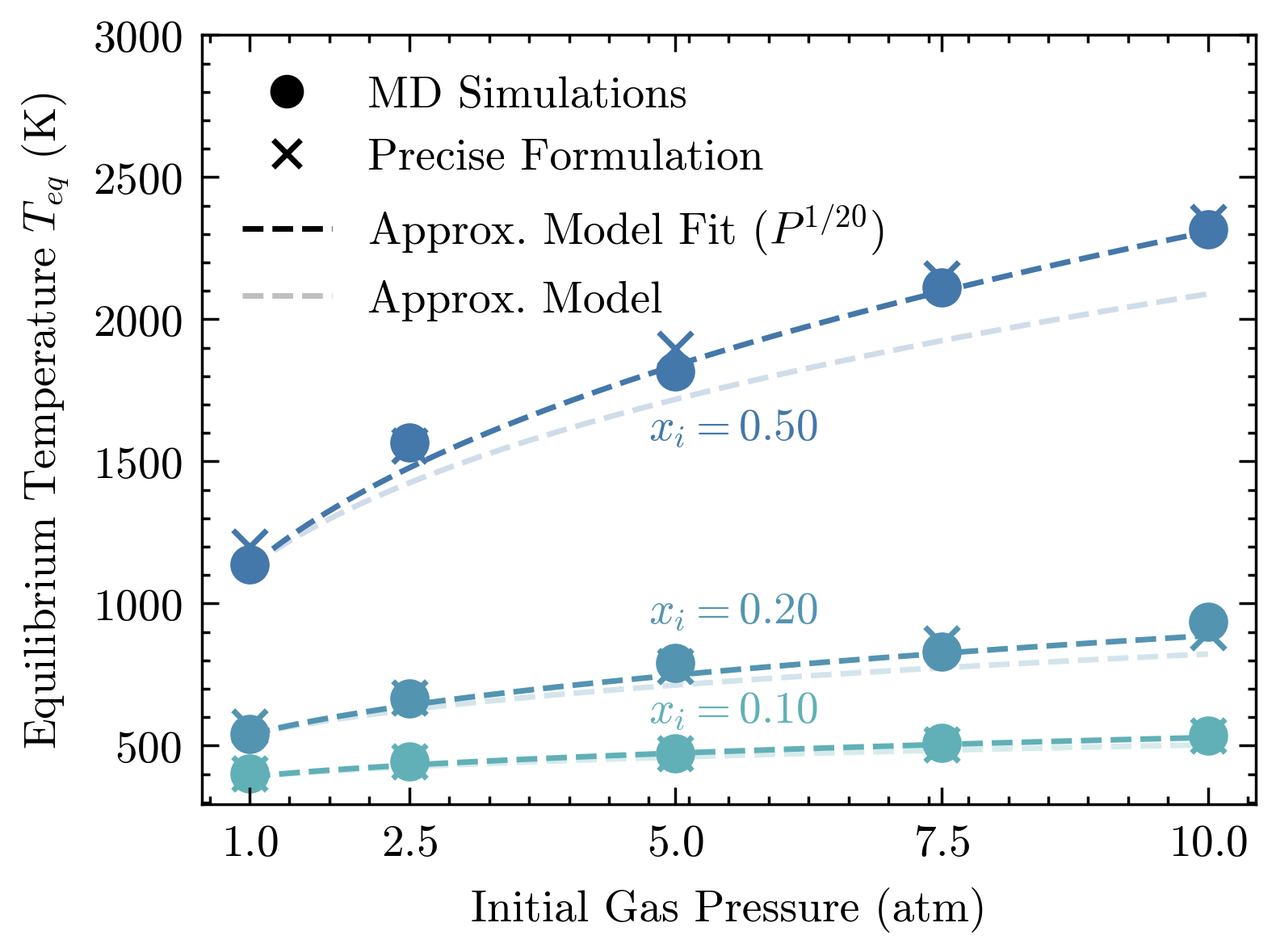}
    \caption{Simulated and predicted temperatures of N$_2$ as a function of the initial gas pressure. The approximate model from equation (\ref{eq:approximate}) must be scaled by $P^{1/20}$ to fit MD data.}
    \label{fig:temp-pressure}
\end{figure}

A higher initial gas pressure leads to a larger ion density and thus, we can expect more heating. MD results, shown in figure \ref{fig:temp-pressure}, confirm this. Additionally, it is shown that the precise formulation from equation~(\ref{gr-eq}) remains accurate, while the approximate model from equation~(\ref{eq:approximate}) requires an increased value for the correction factor $\kappa$ to fit MD data. Scaling $\kappa$ as $P^{1/20}$, where $P$ is the initial gas pressure, gives good agreement. 

\section{Gradual Ionization} \label{gradual}
All simulations discussed thus far have involved an instantaneous ionization process. However, in atmospheric pressure plasmas, ionization processes often occur over the course of several nanoseconds\cite{Tardiveau_2009, Lo_2017}. It has already been shown that, in atomic plasma, the final temperature following disorder-induced heating and subsequent ion-neutral temperature equilibration are unaffected by the rate of ionization~\cite{gradual}. Considering that translational-rotational relaxation occurs on the same timescale as ion-neutral relaxation in these plasmas, we expect this to be the case for molecular plasmas as well. 

\begin{figure}
    \centering
    \includegraphics[width=0.5\textwidth]{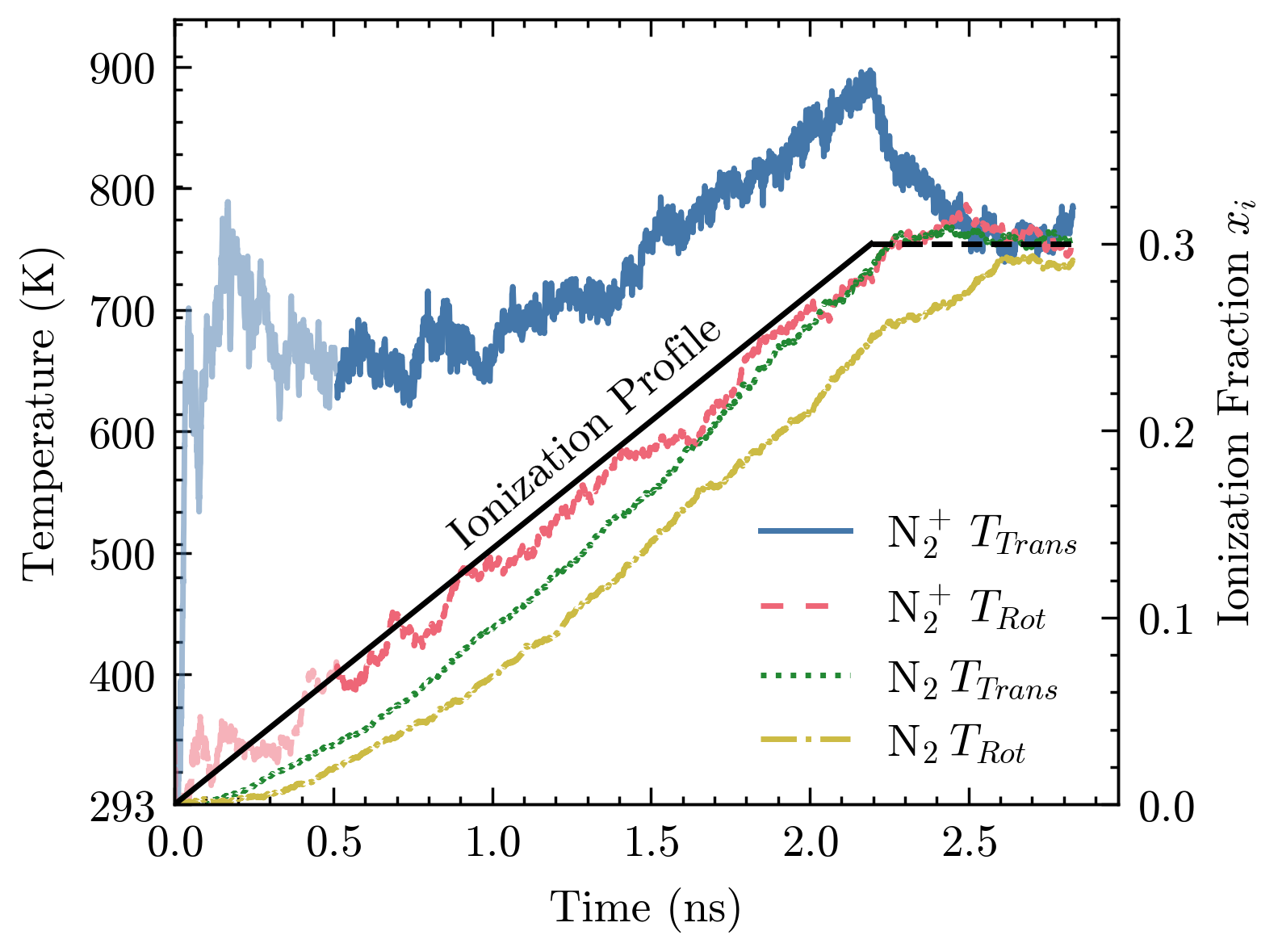}
    \caption{Simulated time-evolution of the temperature components of an N$_2$ plasma subject to a gradual ionization process (2.2 ns) until a 0.30 ionization fraction is reached. Initial ion temperatures are made a lighter color because early in the simulation, there are not enough ions to have a well-defined temperature.}
    \label{fig:gradual}
\end{figure}

To confirm this, an MD simulation was run with a gradual ionization process. The ion density was increased linearly over the course of 2.2 ns until an ionization fraction of 0.30 was reached. The results, shown in figure \ref{fig:gradual}, confirm expectations. The simulation reaches an equilibrium temperature of 752 K, matching the simulations from an instantaneous ionization process to a 2.7\% error. It can also be observed that all degrees of freedom present in the simulation heat at, roughly, the rate of ionization. This is because ion-neutral and translational-rotational relaxation in these plasmas occur on a faster timescale than the ionization process. Finally, it should be noted that early on in the simulation, when the ionization fraction is low, there are not enough ions to have a well-defined temperature. Consequently, the early temperature data has a significant amount of noise.

The relaxation dynamics present in both the instantaneous and gradual ionization process are complex. Strongly coupled ion-ion collisions play a large role in the system's translational-rotational energy exchange, so attempts to model the relaxation with the Boltzmann equation\cite{ferziger1972mathematical} and Parker's model\cite{parker1959rotational} (a model for RT relaxation in neutral gas) fail, as they do not account for strong coupling. Thus, this work motivates the need for a theory of translational-rotational relaxation in a strongly coupled plasma.

\section{Conclusions}

This work examined how plasma and gas heating due to DIH changes with the introduction of molecules and higher pressures. Using MD, it was shown that energy from DIH is spread to rotational degrees of freedom on a nanosecond timescale, leading to a lower equilibrium temperature than in the atomic case. It was also shown that a higher initial gas pressure leads to a larger temperature increase from DIH. Two models were derived that can predict this behavior to a high degree of accuracy. Finally, by simulating a gradual ionization over the course of a few nanoseconds, it was shown that the temperature change due to DIH in molecular plasmas is unaffected by ionization dynamics.  

This work shows that the effects of strong coupling remain significant in molecular plasmas and establish that one must consider all active degrees of freedom in order to accurately model the influence of DIH on the overall heating process. It further motivates the need to develop new modelling techniques that can account for strong coupling in CAPPs, as current models based on the Boltzmann equation fail to do this. These models do not consider DIH and as such, they risk vastly underestimating the temperature. 

\section{Acknowledgements}

The authors thank Dr.~Lucas Beving for suggesting this topic and Dr.~Christopher Moore for helpful conversations on this study. This research was supported by the US Department of Energy under award no.~DE-SC0022201. It was also supported by computational resources and services provided by Advanced Research Computing (ARC), a division of Information and Technology Services (ITS) at the University of Michigan, Ann Arbor. 

\bibliography{references.bib}

\begin{thebibliography}{10}

\bibitem{10.1063/5.0008093}
Thomas von Woedtke, Steffen Emmert, Hans-Robert Metelmann, Stefan Rupf, and Klaus-Dieter Weltmann.
\newblock {Perspectives on cold atmospheric plasma (CAP) applications in medicine}.
\newblock {\em Physics of Plasmas}, 27(7), 2020.
\newblock 070601.

\bibitem{app11083372}
Azadeh Barjasteh, Zohreh Dehghani, Pradeep Lamichhane, Neha Kaushik, Eun~Ha Choi, and Nagendra~Kumar Kaushik.
\newblock Recent progress in applications of non-thermal plasma for water purification, bio-sterilization, and decontamination.
\newblock {\em Applied Sciences}, 11(8), 2021.

\bibitem{https://doi.org/10.1111/jfpp.15070}
Abirami~R. Ganesan, Uma Tiwari, P.~N. Ezhilarasi, and Gaurav Rajauria.
\newblock Application of cold plasma on food matrices: A review on current and future prospects.
\newblock {\em Journal of Food Processing and Preservation}, 45(1):e15070, 2021.

\bibitem{article}
Noala Vicensoto Moreira~Milhan, William Chiappim~Junior, Aline da~Graça~Sampaio, Mariana Vegian, Rodrigo Pessoa, and Cristiane Koga~Ito.
\newblock Applications of plasma-activated water in dentistry: A review.
\newblock {\em International Journal of Molecular Sciences}, 23:4131, 2022.

\bibitem{KUMAR2021100197}
S.P.~Jeevan Kumar, Anjani~Devi Chintagunta, Y.~Mohan Reddy, Loïc Rajjou, Vijay~Kumar Garlapati, Dinesh~K. Agarwal, S.~Rajendra Prasad, and Jesus Simal-Gandara.
\newblock Implications of reactive oxygen and nitrogen species in seed physiology for sustainable crop productivity under changing climate conditions.
\newblock {\em Current Plant Biology}, 26:100197, 2021.

\bibitem{bioengineering10030280}
Som~V. Thomas, Krista Dienger-Stambaugh, Michael Jordan, Yuxin Wang, Jason Hammonds, Paul Spearman, and Donglu Shi.
\newblock Inactivation of sars-cov-2 on surfaces by cold-plasma-generated reactive species.
\newblock {\em Bioengineering}, 10(3), 2023.

\bibitem{app11114809}
Mária Domonkos, Petra Tichá, Jan Trejbal, and Pavel Demo.
\newblock Applications of cold atmospheric pressure plasma technology in medicine, agriculture and food industry.
\newblock {\em Applied Sciences}, 11(11), 2021.

\bibitem{Neyts_2014}
Erik~C Neyts, Maksudbek Yusupov, Christof~C Verlackt, and Annemie Bogaerts.
\newblock Computer simulations of plasma–biomolecule and plasma–tissue interactions for a better insight in plasma medicine.
\newblock {\em Journal of Physics D: Applied Physics}, 47(29):293001, 2014.

\bibitem{doi:10.1080/10408398.2018.1564731}
Kexin Zhang, Camila~A. Perussello, Vladimir Milosavljević, P.~J. Cullen, Da-Wen Sun, and Brijesh~K. Tiwari.
\newblock Diagnostics of plasma reactive species and induced chemistry of plasma treated foods.
\newblock {\em Critical Reviews in Food Science and Nutrition}, 59(5):812--825, 2019.
\newblock PMID: 30676057.

\bibitem{PMID:30289686}
Yury Gorbanev, Angela Privat-Maldonado, and Annemie Bogaerts.
\newblock Analysis of short-lived reactive species in plasma-air-water systems: The dos and the do nots.
\newblock {\em Analytical chemistry}, 90(22):13151—13158, 2018.

\bibitem{Adamovich_2022}
I~Adamovich, S~Agarwal, E~Ahedo, L~L Alves, S~Baalrud, N~Babaeva, A~Bogaerts, A~Bourdon, P~J Bruggeman, C~Canal, E~H Choi, S~Coulombe, Z~Donkó, D~B Graves, S~Hamaguchi, D~Hegemann, M~Hori, H-H Kim, G~M~W Kroesen, M~J Kushner, A~Laricchiuta, X~Li, T~E Magin, S~Mededovic Thagard, V~Miller, A~B Murphy, G~S Oehrlein, N~Puac, R~M Sankaran, S~Samukawa, M~Shiratani, M~Šimek, N~Tarasenko, K~Terashima, E~Thomas Jr, J~Trieschmann, S~Tsikata, M~M Turner, I~J van~der Walt, M~C~M van~de Sanden, and T~von Woedtke.
\newblock The 2022 plasma roadmap: low temperature plasma science and technology.
\newblock {\em Journal of Physics D: Applied Physics}, 55(37):373001, 2022.

\bibitem{Acciarri_2022}
M~D Acciarri, C~Moore, and S~D Baalrud.
\newblock Strong coulomb coupling influences ion and neutral temperatures in atmospheric pressure plasmas.
\newblock {\em Plasma Sources Science and Technology}, 31(12):125005, 2022.

\bibitem{Lotfy2019}
Khaled Lotfy, Nadi~Awad Al-Harbi, and Hany Abd El-Raheem.
\newblock Cold atmospheric pressure nitrogen plasma jet for enhancement germination of wheat seeds.
\newblock {\em Plasma Chemistry and Plasma Processing}, 39(4):897--912, 2019.

\bibitem{Iuchi2018-hs}
Katsuya Iuchi, Yukina Morisada, Yuri Yoshino, Takahiro Himuro, Yoji Saito, Tomoyuki Murakami, and Hisashi Hisatomi.
\newblock Cold atmospheric-pressure nitrogen plasma induces the production of reactive nitrogen species and cell death by increasing intracellular calcium in {HEK293T} cells.
\newblock {\em Arch Biochem Biophys}, 654:136--145, 2018.

\bibitem{10.1063/1.4884787}
Aijun Yang, Dingxin Liu, Mingzhe Rong, Xiaohua Wang, and Michael~G. Kong.
\newblock {A dominant role of oxygen additive on cold atmospheric-pressure He + O2 plasmas}.
\newblock {\em Physics of Plasmas}, 21(8), 2014.
\newblock 083501.

\bibitem{Aleksandrov11}
Andrey Starikovskiy and Nickolay Aleksandrov.
\newblock Plasma-assisted ignition and combustion.
\newblock In Max Mulder, editor, {\em Aeronautics and Astronautics}, chapter~12. IntechOpen, Rijeka, 2011.

\bibitem{Popov_2011}
N~A Popov.
\newblock Fast gas heating in a nitrogen–oxygen discharge plasma: I. kinetic mechanism.
\newblock {\em Journal of Physics D: Applied Physics}, 44(28):285201, 2011.

\bibitem{CHENG2022111990}
L.~Cheng, N.~Barleon, B.~Cuenot, O.~Vermorel, and A.~Bourdon.
\newblock Plasma assisted combustion of methane-air mixtures: Validation and reduction.
\newblock {\em Combustion and Flame}, 240:111990, 2022.

\bibitem{Pokrovskiy_2022}
G~V Pokrovskiy, N~A Popov, and S~M Starikovskaia.
\newblock Fast gas heating and kinetics of electronically excited states in a nanosecond capillary discharge in co2.
\newblock {\em Plasma Sources Science and Technology}, 31(3):035010, 2022.

\bibitem{POPOV2022100928}
N.A. Popov and S.M. Starikovskaia.
\newblock Relaxation of electronic excitation in nitrogen/oxygen and fuel/air mixtures: fast gas heating in plasma-assisted ignition and flame stabilization.
\newblock {\em Progress in Energy and Combustion Science}, 91:100928, 2022.

\bibitem{Lo_2017}
A~Lo, A~Cessou, C~Lacour, B~Lecordier, P~Boubert, D~A Xu, C~O Laux, and P~Vervisch.
\newblock Streamer-to-spark transition initiated by a nanosecond overvoltage pulsed discharge in air.
\newblock {\em Plasma Sources Science and Technology}, 26(4):045012, 2017.

\bibitem{Minesi_2020}
N~Minesi, S~Stepanyan, P~Mariotto, G~D Stancu, and C~O Laux.
\newblock Fully ionized nanosecond discharges in air: the thermal spark.
\newblock {\em Plasma Sources Science and Technology}, 29(8):085003, 2020.

\bibitem{vanderHorst_2012}
R~M van~der Horst, T~Verreycken, E~M van Veldhuizen, and P~J Bruggeman.
\newblock Time-resolved optical emission spectroscopy of nanosecond pulsed discharges in atmospheric-pressure n2 and n2/h2o mixtures.
\newblock {\em Journal of Physics D: Applied Physics}, 45(34):345201, 2012.

\bibitem{Benilov_2008}
M~S Benilov.
\newblock Understanding and modelling plasma–electrode interaction in high-pressure arc discharges: a review.
\newblock {\em Journal of Physics D: Applied Physics}, 41(14):144001, jul 2008.

\bibitem{bergeson}
S.~D. Bergeson, M.~Lyon, J.~B. Peatross, N.~Harrison, D.~Crunkleton, J.~Wilson, S.~Rupper, A.~Diaw, and M.~S. Murillo.
\newblock {A neutral strongly coupled laser-produced plasma by strong-field ionization in a gas jet}.
\newblock {\em AIP Conference Proceedings}, 1668(1):040001, 06 2015.

\bibitem{Bataller_2014}
A.~Bataller, G.~R. Plateau, B.~Kappus, and S.~Putterman.
\newblock Blackbody emission from laser breakdown in high-pressure gases.
\newblock {\em Phys. Rev. Lett.}, 113:075001, Aug 2014.

\bibitem{Bataller:19}
Alexander Bataller, Alexandra Latshaw, John~P. Koulakis, and Seth Putterman.
\newblock Dynamics of strongly coupled two-component plasma via ultrafast spectroscopy.
\newblock {\em Opt. Lett.}, 44(23):5832--5835, Dec 2019.

\bibitem{gradual}
M~D Acciarri, C~Moore, and S~D Baalrud.
\newblock Disorder-induced heating as a mechanism for fast neutral gas heating in atmospheric pressure plasmas.
\newblock {\em Plasma Sources Science and Technology}, 33(2):02LT02, 2024.

\bibitem{PhysRevLett.87.115003}
M.~S. Murillo.
\newblock Using fermi statistics to create strongly coupled ion plasmas in atom traps.
\newblock {\em Phys. Rev. Lett.}, 87:115003, 2001.

\bibitem{PhysRevLett.88.065003}
S.~G. Kuzmin and T.~M. O'Neil.
\newblock Numerical simulation of ultracold plasmas: How rapid intrinsic heating limits the development of correlation.
\newblock {\em Phys. Rev. Lett.}, 88:065003, 2002.

\bibitem{PhysRevLett.83.4776}
T.~C. Killian, S.~Kulin, S.~D. Bergeson, L.~A. Orozco, C.~Orzel, and S.~L. Rolston.
\newblock Creation of an ultracold neutral plasma.
\newblock {\em Phys. Rev. Lett.}, 83:4776--4779, 1999.

\bibitem{PhysRevE.94.021201}
D.~Murphy and B.~M. Sparkes.
\newblock Disorder-induced heating of ultracold neutral plasmas created from atoms in partially filled optical lattices.
\newblock {\em Phys. Rev. E}, 94:021201, 2016.

\bibitem{dihexperiment}
Y.~C. Chen, C.~E. Simien, S.~Laha, P.~Gupta, Y.~N. Martinez, P.~G. Mickelson, S.~B. Nagel, and T.~C. Killian.
\newblock Electron screening and kinetic-energy oscillations in a strongly coupled plasma.
\newblock {\em Phys. Rev. Lett.}, 93:265003, Dec 2004.

\bibitem{Foster_Fetsch_Fisch_2023}
Thomas~E. Foster, Henry Fetsch, and Nathaniel~J. Fisch.
\newblock Fast correlation heating in moderately coupled electron–ion plasmas.
\newblock {\em Journal of Plasma Physics}, 89(5):905890506, 2023.

\bibitem{LAMMPS}
A.~P. Thompson, H.~M. Aktulga, R.~Berger, D.~S. Bolintineanu, W.~M. Brown, P.~S. Crozier, P.~J. in~'t Veld, A.~Kohlmeyer, S.~G. Moore, T.~D. Nguyen, R.~Shan, M.~J. Stevens, J.~Tranchida, C.~Trott, and S.~J. Plimpton.
\newblock {LAMMPS} - a flexible simulation tool for particle-based materials modeling at the atomic, meso, and continuum scales.
\newblock {\em Comp. Phys. Comm.}, 271:108171, 2022.

\bibitem{lide2004crc}
David~R Lide.
\newblock {\em CRC handbook of chemistry and physics}, volume~85.
\newblock CRC press, 2004.

\bibitem{mcquarrie1997physical}
Donald~Allan McQuarrie and John~Douglas Simon.
\newblock {\em Physical chemistry: a molecular approach}, volume~1.
\newblock University science books Sausalito, CA, 1997.

\bibitem{10.1063/5.0021993}
Wei Yang, Qianhong Zhou, Qiang Sun, Zhiwei Dong, and Eryan Yan.
\newblock {Vibrational–translational relaxation in nitrogen discharge plasmas: Master equation modeling and Landau–Teller model revisited}.
\newblock {\em AIP Advances}, 10(10):105311, 2020.

\bibitem{BAUS19801}
Marc Baus and Jean-Pierre Hansen.
\newblock Statistical mechanics of simple coulomb systems.
\newblock {\em Physics Reports}, 59(1):1--94, 1980.

\bibitem{doi:10.1021/jp952026w}
Mikhail~N. Glukhovtsev and Sergei Laiter.
\newblock Thermochemistry of tetrazete and tetraazatetrahedrane: A high-level computational study.
\newblock {\em The Journal of Physical Chemistry}, 100(5):1569--1577, 1996.

\bibitem{hansen2006theory}
J.P. Hansen and I.R. McDonald.
\newblock {\em Theory of Simple Liquids}.
\newblock Elsevier Science, 2006.

\bibitem{Gould2010-vd}
Harvey Gould and Jan Tobochnik.
\newblock {\em Statistical and thermal physics}.
\newblock Princeton University Press, Princeton, NJ, 2010.

\bibitem{STELL1963517}
George Stell.
\newblock The percus-yevick equation for the radial distribution function of a fluid.
\newblock {\em Physica}, 29(5):517--534, 1963.

\bibitem{10.1143/PTP.20.920}
Tohru Morita.
\newblock {Theory of Classical Fluids: Hyper-Netted Chain Approximation, I: Formulation for a One-Component System}.
\newblock {\em Progress of Theoretical Physics}, 20(6):920--938, 1958.

\bibitem{Tardiveau_2009}
P~Tardiveau, N~Moreau, S~Bentaleb, C~Postel, and S~Pasquiers.
\newblock Diffuse mode and diffuse-to-filamentary transition in a high pressure nanosecond scale corona discharge under high voltage.
\newblock {\em Journal of Physics D: Applied Physics}, 42(17):175202, 2009.

\bibitem{ferziger1972mathematical}
J.H. Ferziger and H.G. Kaper.
\newblock {\em Mathematical Theory of Transport Processes in Gases}.
\newblock North-Holland Publishing Company, 1972.

\bibitem{parker1959rotational}
JG~Parker.
\newblock Rotational and vibrational relaxation in diatomic gases.
\newblock {\em The Physics of Fluids}, 2(4):449--462, 1959.

\end{thebibliography}
\end{document}